\documentclass[aps,pra,twocolumn,nopacs,floatfix]{revtex4}
\usepackage{epsfig}
\usepackage{graphicx}
\usepackage{dcolumn}
\usepackage{amsthm,amsmath}
\usepackage{comment}
\usepackage[colorlinks]{hyperref}
\usepackage{xcolor}
\usepackage{longtable}
\usepackage{amsmath}

\begin{document}

\title{\bf{Accurate determination of quadrupole polarizabilities of the excited states of alkali-metal atoms}}

\author{$^1$Harpreet Kaur}
\author{$^1$Sukhjit Singh}
\author{$^1$Bindiya Arora}
\email{bindiya.phy@gndu.ac.in}
\author{$^2$B. K. Sahoo}

\affiliation{$^1$Department of Physics, Guru Nanak Dev University, Amritsar, Punjab 143005, India}
\affiliation{$^2$Atomic, Molecular and Optical Physics Division, Physical Research Laboratory, Navrangpura, Ahmedabad-380009, India}

\begin{abstract}
The scalar and tensor components of the electric quadrupole (E2) polarizabilities of the first two excited states of all the alkali-metal atoms 
are determined. To validate the calculations, we have evaluated the ground state E2 polarizabilities of these atoms and compared them with the 
literature values. We could not find the ground state E2  polarizability value for Fr in the literature to compare with our result. The dominant parts of these quantities are estimated by combining the precisely calculated E2 transition matrix elements of 
many low-lying transitions with the experimental energies, while the other contributions are estimated using lower-order methods. Our 
estimated values for the ground states of the above atoms are in good agreement with the literature values suggesting that our estimated E2 
polarizabilities for the excited states of the alkali atoms, which were not known earlier except for the Li atom, are also quite accurate. These 
reported E2 polarizabilities could be useful in guiding many precision measurements in the alkali atoms.
\end{abstract}

\maketitle

\section{Introduction}\label{sec1}

Studies of electric polarizabilities of atoms, molecules and clusters are highly demanding for both the experimental and theoretical perspectives 
\cite{maroulis2006atoms,PhysRevA.64.025201}. High precision values of polarizabilities of these systems are very useful in several branches of 
science~\cite{iskrenova2007high,yan2004molecular,doi:10.1146/annurev.physchem.48.1.213}. Some of the examples of prominent applications of
electric polarizabilities in atomic systems, which are of present interest, include the optical atomic clock measurements \cite{udem2002optical}, 
discrete symmetry violations~\cite{roberts2015parity}, condensates of dilute atomic gases \cite{doi:10.1126/science.269.5221.198}, etc.~\cite{karshenboim2004astrophysics,
blaga2009strong}. Atoms are spherically symmetric, but upon the influence of stray electric fields result in multi-order shifts in energy levels~\cite{bonin}. The 
interaction between any system and an electric field is predominately treated in the framework of electric dipole (E1) approximation \cite{bonin}.
However, higher-order contributions from the interaction of quadrupole operator with external electric field gradient may become significant for 
some of the applications that aim to achieve ultra-precision measurements~\cite{shiratori2012theory,doi:10.1063/1.1578052,HUTSON1986L775,
doi:10.1063/1.4940397,PhysRevLett.112.106101,zhang2013electric}. The first-order shift due to electric 
quadrupole (E2) interaction renders E2 moment that is generally zero for atomic states with angular momentum $J<3/2$; otherwise they can also 
offer non-vanishing contributions~\cite{PhysRevA.93.012503}. The second-order shift gives finite E2 polarizability and plays a dominant part after 
E1 polarizabilities and E2  moments~\cite{PhysRevA.98.013406}. Particularly, E2 polarizabilities arising due to contribution from the forbidden 
transitions, might play a significant role in deducing one of the dominant environment induced blackbody radiation (BBR) shifts in order to reach 
the required accuracy level below $10^{-19}$ in the atomic clocks \cite{PhysRevA.74.020502,PhysRevA.88.013405,PhysRevA.103.013109}. 

Alkali metal atoms are desired for many experimental studies as they form well controlled and characterized systems. Owing to their easily laser 
accessible level structures, they are used for vital applications such as atomic clocks~\cite{10.1007/978-3-319-67380-6_52, bauch2003caesium}, 
scattering phenomena~\cite{PhysRevA.67.013814}, quantum computation \cite{Wu_2021}, quantum sensing~\cite{Fan_2015}, cold collision
\cite{PhysRevA.44.4464}, long-range interactions \cite{Singer_2005}, etc. Among alkali atoms, the Rb and Cs atoms 
have been considered as the best candidates for microwave clocks~\cite{5422509,PhysRevLett.101.220801}, whereas the Fr atom has attracted a lot of 
attention as a candidate for studying fundamental symmetry tests~\cite{PhysRevA.90.052502, stancari2007francium, FrPNC,FrEDM2}. 
Since there is a similarity between the energy level spacing between Rb and Fr, Fr can also be laser cooled and trapped using available 
lasers~\cite{2011}. This is what, besides being the heaviest alkali atom, makes it favorite for probing new physics beyond the 
Standard Model of elementary particles~\cite{FrEDM1}. 

Precise determination of E2 polarizabilities for alkali atoms have been given a very little attention, especially for the excited states. 
Compared to the E1 polarizabilities, it is strenuous to measure the E2 polarizabilities in atomic systems due to  their extremely weak
contributions to the energy level shifts caused by the spatial gradient of electric fields. This is why accurate theoretical calculations 
of these quantities are very crucial.  While a number of theoretical studies exist for the ground state E2 polarizabilities of the alkali atoms~\cite{doi:10.1063/1.473089,PhysRevA.74.020502,jiang2015effective,PhysRevA.77.022510,PhysRevA.78.052504,PhysRevA.83.052508,PhysRevA.94.012505}
(except for Fr), very limited studies have been conducted for the excited states~\cite{PhysRevA.78.012515}. Since the D1 and 
D2 lines of the alkali atoms directly participate in the laser cooling process of alkali atoms, accurate knowledge of the E2 polarizabilities 
for the first two excited states of alkali atoms are quite useful. Furthermore, accurate knowledge of quadrupole polarizabilities are essential 
for estimating dispersion potentials among atomic systems~\cite{tao2012accurate,PhysRevA.81.052507}. The importance of polarizabilities of excited 
states of atoms was demonstrated by Zhu \textit{et al.} in the studies of long-range interactions of the alkali-metal atoms in their 
ground and excited states
with helium atom for astrophysical applications~\cite{PhysRevA.70.032722}. Their reliable values are also crucial in order to describe the 
the van der Waal atom-surface potentials~\cite{hutson1986quadrupolar,PhysRevA.81.052507,laliotis2021atom}. Accurate values of quadrupole polarizabilities are 
required to construct the scattering potentials in the ultracold physics and determining scattering cross-sections of electrons or positrons
from an atomic system \cite{C7CP02127E,doi:10.1080/00107514.2013.854618}. Precise calculations of polarizabilities of heavier atomic systems 
depend upon the potential of the many-body method used to account for the relativistic and electron correlation effects \cite{doi:10.1063/1.466280,
KELLO1996383,PhysRevA.60.2822,PhysRevA.70.062501}. 

Previously, the static E2 polarizabilities for the ground states of the alkali atoms have been calculated using simple analytic wave functions by 
Patil \textit{et al.}~\cite{doi:10.1063/1.473089} and semi-empirical calculations by Jiang \textit{et al.}~\cite{jiang2015effective}. Combining 
the relativistic many-body perturbation theory (RMBPT) and random phase approximations (RPA), the ground state E2 polarizabilities of the 
alkali-metals have been evaluated by Porsev and Derevianko~\cite{PhysRevA.74.020502}. Safronova \textit{et al.} have calculated high-precision 
spectroscopic properties including E2 polarizabilities of the ground states of Li, K, Rb and Cs using linearized coupled-cluster method~\cite{PhysRevA.77.022510,PhysRevA.78.052504,PhysRevA.83.052508,PhysRevA.94.012505}. The static E2 polarizabilities of the ground 
state and a few low-lying excited states of Li have been evaluated by Wansbeek \textit{et al.} by adopting relativistic coupled-cluster method 
in fully \textit{ab initio} procedure~\cite{PhysRevA.78.012515}.

In the present work, we conduct extensive calculations of many E2 matrix elements of the transitions of alkali-metal atoms using the relativistic 
all-order (AO) method that predominantly contribute to the determination of E2 polarizabilities. We provide both the scalar and tensor components 
of the E2 polarizabilities of the excited $nP_{3/2}$ states, with the ground state principal quantum number $n$, along with the 
scalar E2 polarizabilities of the ground states $nS_{1/2}$ and excited state $nP_{1/2}$, of the considered atoms. The accuracy of these quantities are estimated by comparing 
the E2 matrix elements and polarizability values of the ground states from the previous works. The bifurcation of the paper is as follows: 
Sec.~\ref{sec2} includes a brief theory on E2 polarizability. Sec.~\ref{sec3} consists of methods of evaluation of wave functions and E2 matrix 
elements in the framework of relativistic all-order approach. The E2 polarizability results along with their uncertainties have been given and 
discussed in Sec.~\ref{sec4}. Finally, we have concluded our work in Sec.~\ref{sec5}.

\section{Theory}\label{sec2}

When an atom is placed in a static electric field, it experiences shifts in the energy levels which can be conveniently expressed in terms of
electric multipole effects. In particular, the perturbation interaction Hamiltonian, $H_{int}^{q}= {\bf Q} \cdot {\vec \nabla} \cal{E}$ with 
quadruploe operator ${\bf Q}=\sum_i q_i $, due to interaction of quadrupole effect with the gradient of an electric field (${\vec \nabla} \cal{E}$) gives
second-order energy shift in the energy level of an atom in state $|\Psi_{n}\rangle$, is given by 
\begin{equation}
\label{Eshift}
\Delta E_{n}^{q(2)}=\sum_{k \neq n} \left[\frac{ (q*)_{nk}(q)_{kn}}{\delta E_{nk}}\right].
\end{equation}
where $(q)_{nk}=\langle\Psi_{n}|H_{int}^{q}|\Psi_{k}\rangle$ with $k$ denoting the index for the intermediate states that are permitted by the 
quadrupole selection rules and $\delta E_{nk} = E_n - E_k $ with $E_{i=n,k}$'s are the energies of the corresponding states. The quadrupole 
moments of the $nP_{3/2}$ states, which can be used to estimate the first-order effects, in the considered systems have been determined accurately 
earlier \cite{STONE20161}. For the computational simplicity, $\Delta E_{n}^{q(2)}$ for linearly polarized light with 
polarization vector along the quantization axis can be expressed as~\cite{PhysRevA.86.022506} 
\begin{eqnarray}
\label{eq5}
\Delta E_{n}^{q(2)} = - \frac{1}{8} \alpha_{n}^q {(\nabla \cal E)}^2
\end{eqnarray}
with 
\begin{eqnarray}
\alpha_{n}^q &=& \left[ \alpha_{n}^{q(0)} -  \alpha_{n}^{q(2)} \frac{3M_{J_n}^2-{J_n}({J_n}+1)}{{J_n}(2{J_n}-1)} \right. \nonumber \\
& & \left. - 3 \alpha_{n}^{q(4)} (5M_n^2-J_n^2-2J_n) \right. \nonumber \\
& & \left. \times \frac{(5M_n^2+1-J_n^2)-10M_n^2(4M_n^2-1)}{J_n(J_n-1)(2J_n-1)(2J_n-3)}  \right],
\end{eqnarray}
where $M_{J_n}$ is the magnetic quantum number. Here $\alpha_{n}^q$ is the total quadrupole polarizability, which is given in terms of $M_{J_n}$
independent quantities as $\alpha_{n}^{q(0)}$, $\alpha_{n}^{q(2)}$ and $\alpha_{n}^{q(4)}$ -- referred to as the scalar, tensor of rank 2 and
tensor of rank 4 components, respectively~\cite{PhysRevA.98.013406,itano2000external}. It clearly shows that for $J_n=1/2$, contributions from 
both $\alpha_{n}^{q(2)}$ and $\alpha_{n}^{q(4)}$ to $\alpha_{n}^q$ vanish; otherwise they will contribute. Similarly, $\alpha_{nq}^{(4)}$ is 
non-zero when $J_n>3/2$. Since we consider states with $J_n={1/2}$ and $J_n={3/2}$ in the present work, contributions from $\alpha_{n}^{q(4)}$ 
become irrelevant. Expressions for the $M_{J_n}$ independent $\alpha_{n}^{q(0)}$ and $\alpha_{n}^{q(2)}$ are given by~\cite{PhysRevA.98.013406}
\begin{eqnarray}
\label{eq6}
\alpha_{n}^{q(0)}& = & - 2\sum_{k \neq n} W_{n}^{q(0)} \left[\frac{|\langle \psi_n||\textbf{Q}||\psi_k \rangle|^2}{\delta E_{nk}}\right] \label{scalar}
\end{eqnarray}
and
\begin{eqnarray}
\alpha_{n}^{q(2)}& = & - 2\sum_{k \neq n} W_{n,k}^{q(2)} \left[\frac{|\langle \psi_n||\textbf{Q}||\psi_k \rangle|^2 }{\delta E_{nk}}\right],\label{tensor}
\end{eqnarray}
where the factors $W_{n}^{q(0)}$ and $W_{n,k}^{q(2)}$ are given by
\begin{equation}
\label{eq8}
W_{n}^{q(0)} = \frac{1}{5(2J_n+1)}
\end{equation}
and
\begin{eqnarray}
\label{eq9}
W_{n,k}^{q(2)} &=& \sqrt{\frac{10J_n(2J_n-1)}{7(J_n+1)(2J_n+1)(2J_n+3)}}
                                \nonumber \\
& &\times (-1)^{J_n+J_k+1}  \left\{ \begin{array}{ccc}
                                            J_n & 2 & J_n\\
                                            2 & J_k & 2
                                           \end{array}\right\} 
\end{eqnarray}
with $ \left\{ \begin{array}{ccc}
                                            J_n & 2 & J_n\\
                                            2 & J_k & 2
                                           \end{array}\right\}$
as the Wigner angular momentum coupling 6-j symbol.

\section{Method of evaluation}\label{sec3}

The procedure to determine wave functions of the ground and intermediate states of alkali atoms using relativistic AO method are already
presented in Ref.~\cite{PhysRevA.75.042515}. In brief, using Dirac-Fock (DF) method, the electronic configuration of alkali atoms are divided into a closed-core and a valence orbital in order to obtain 
the mean-field wave function of the respective closed-shell ($|0_{c}\rangle$). Further, the mean-field 
wave functions of the atomic states are obtained by appending the respective valence orbital $v$ as 
\begin{equation}
|\phi_v\rangle = a_v^{\dagger}|0_{c}\rangle .
\end{equation}
To obtain the DF orbitals, we use a set of 50 B-splines of order $k = 11$ for each angular momentum. The basis set orbitals are constrained to a 
large spherical cavity of a radius $R = 220$ a.u.. 

Contribution to the evaluation of a matrix element can be divided into core, core-valence and
valence contributions as described in Ref.~\cite{PhysRevA.91.012705} which in turn divide the scalar and tensor components of polarizability from Eqs. (\ref{scalar}) and (\ref{tensor}) into respective contributions of polarizability given as 
\begin{equation}
\label{eq11}
\alpha_{n}^{q(t=0,2)} = \alpha_{n,c}^{q(t=0,2)} + \alpha_{n,vc}^{q(t=0,2)} + \alpha_{n,v}^{q(t=0,2)},
\end{equation}
where superscript $t$ denotes the scalar ($t=0$) and tensor ($t=2$) components of polarizability, and subscripts $c$, $vc$ and $v$ denote contributions from core, core-valence and valence correlations respectively. It can be noted that 
$\alpha_{n,c}^{q(0)}$ is same for all atomic states as they have a common closed-core while $\alpha_{n,c}^{q(2)}$ is zero. Compared to 
$\alpha_{n,v}^{q(t=0,2)}$, magnitudes of $\alpha_{n,c}^{q(t=0,2)}$ and $\alpha_{n,vc}^{q(t=0,2)}$ are typically much smaller. These dominating 
valence contributions need to be estimated precisely for accurate determination of E2 polarizabilities. The $\alpha_{n,v}^{q(t=0,2)}$ 
contributions are evaluated by
\begin{eqnarray}\label{mval}
\alpha_{n,v}^{q(t=0,2)}= -2 \sum_{k > N_c,k\neq n} W_{n}^{q(t=0,2)} \left[ \frac{|\langle\psi_n||\textbf{Q}||\psi_k\rangle|^2}
{\delta E_{nk}}\right] ,
\end{eqnarray}
where the sum is restricted by the number of core orbitals $N_c$ to exclude their contributions. We calculate many E2 matrix elements up to $k \leq I$ states that 
contribute significantly to the above quantity using our relativistic AO method and use experimental energies from the National Institute of 
Science and Technology (NIST) database~\cite{NIST_ASD}. These contributions are referred as main part and are denoted by
$\alpha_{n,v(M)}^{q(t=0,2)}$ in the present work. To evaluate the E2 matrix elements for the main part, atomic wave functions $|\psi_v\rangle$, with $v$ denoting different 
valence orbitals, are expressed in the singles and doubles approximated (SD) all-order method as \cite{Safronova_1999} 
\begin{eqnarray}\label{eq12}
& &|\psi_v\rangle_{SD} = \left[1+ \sum_{ma}\rho_{ma}a_m^\dagger a_a +\frac{1}{2}\sum_{mlab}\rho_{mlab} a_m^\dagger a_l^\dagger a_b a_a\right.
\nonumber \\
& & \left.+ \sum_{m\neq v} \rho_{mv} a_m^\dagger a_v + \sum_{mla} \rho_{mlva}a_m^\dagger a_l^\dagger a_a a_v \right]|\phi_v\rangle ,
\end{eqnarray}						
where $a^\dagger$ and $a$ represent second quantization creation and annihilation operators, respectively, whereas excitation coefficients are 
denoted by $\rho$. The subscripts $m, l, r$ and $a, b, c$ refer to the virtual and core orbitals, respectively. $\rho_{ma}$ and $\rho_{mv}$ are the single whereas 
$\rho_{mlab}$ and $\rho_{mlva}$ are the double excitation coefficients. In addition to this, we also evaluated wave functions that includes the missing third-order terms, by adding the two triple-excitation coefficients - $\rho_{mlrabc}^{pert}$ and 
$\rho_{mlrvab}^{pert}$ perturbatively in the SD wave function solving equation
(SDpT) by defining as follows~\cite{Safronova_1999} 
\begin{eqnarray}\label{eq13}		
|\psi_v\rangle_{SDpT} &=& |\psi_v\rangle_{SD} + \left[ \frac{1}{18} \sum_{mlrabc} \rho_{mlrabc}^{pert} a_m^\dagger a_l^\dagger a_r^\dagger a_c a_b a_a \right.
						\nonumber\\
& & \left. + \frac{1}{6} \sum_{mlrab} \rho_{mlrvab}^{pert} a_m^\dagger a_l^\dagger a_r^\dagger a_b a_a a_v \right]|\phi_v \rangle.   
	 \end{eqnarray}
After obtaining wave functions of the considered states of alkali-metal atoms, we determine E2 matrix elements using the following expression~\cite{PhysRevA.40.2233}
\begin{eqnarray}\label{13}
	Q_{vk} = \frac{\langle\psi_v|\textbf{Q}|\psi_k\rangle}{\sqrt{\langle\psi_v|\psi_v\rangle \langle\psi_k|\psi_k\rangle}}.
	\end{eqnarray}	
In order to estimate contributions due to the neglected physical effects, we scale the wave functions (through the amplitudes of the excitation 
coefficients) to match the calculated energies with their experimental values \cite{PhysRevA.78.022514}; i.e.
\begin{equation}
\rho'_{mv} = \rho_{mv}\frac{\delta E_v^{expt}}{\delta E_v^{theory}},
\end{equation} 	
where $\delta E_v^{expt}$ are the energy differences between the experimental and DF values, and $\delta E_v^{theory}$ are the energy differences
between the experimental results and our final calculations. Then, the E2 matrix elements are reevaluated using the modified excitation amplitudes. 
By analysing the differences between the {\it ab initio} values and the scaled values of the E2 matrix elements, we quote the uncertainties to 
the E2 matrix elements. 

Contributions from the remaining excited states including continuum for valence polarizability are estimated separately using the DF method which 
are referred as tail part of the valence contribution ($\alpha_{n,v(T)}^{q(t=0,2)}$) and are evaluated using the relation  
\begin{equation}\label{Tval}
\alpha_{n,v(T)}^{q(t=0,2)}=-2 \sum_{k > I} W_{n}^{q(t=0,2)} \left[\frac{|\langle\phi_n||{\bf Q}||\phi_k\rangle|^2}{\delta \epsilon_{nk}}\right],
\end{equation}
with $\delta \epsilon_{nk} = \epsilon_n - \epsilon_k$ for the DF energies $\epsilon_i$ and the sum $k>I$ corresponding to the excited states
whose matrix elements are not accounted earlier. The valence-core contributions ($\alpha_{n,vc}^{q(t=0)}$) are estimated using the DF method. To
estimate the core contribution ($\alpha_{n,c}^{q(t=0)}$), however, we have used the following formula~\cite{PhysRevA.77.062516,PhysRevA.90.022511}
\begin{equation}
\alpha_{n,c}^{q(t=0)} = \langle\phi_{n}|\textbf{Q}|\psi_{n}^{(1)}\rangle 
\end{equation}
where $|\psi_{n}^{(1)}\rangle = \sum_{k \neq n} c_{k} |\phi_k \rangle \frac{\langle\phi_k|\textbf{Q}|\phi_n\rangle}{\epsilon_k - \epsilon_n}$ is 
the first-order perturbed wave function due to application of $\textbf{Q}$ operator on ground state $|\psi_n\rangle$ with $c_{k}$ is a 
coefficient containing all-order core-polarization effects due to the residual Coulomb interactions. We have obtained 
$|\psi_{n}^{(1)}\rangle$ in the random phase approximation (RPA) as described in~\cite{PhysRevA.90.022511}.

\begin{table*}[t]
\caption{\label{table1} Contributions to the ground state quadrupole polarizabilities (in a.u.) of the Li, Na, K, Rb, Cs and Fr atoms. Uncertainties to the 
estimates values are quoted in the parentheses. Final results are compared with the previously available values.}
	\begin{center}
\begin{tabular}{|ccccccc|}
\hline
\hline

& Li & Na & K & Rb & Cs & Fr \\

 $\alpha_{n,v(M)}^{q(0)}$   & 1310 & 1773 & 4866 & 6209 & 9670 &	 7909  \\
 
  $\alpha_{n,v(T)}^{q(0)}$ & 114(3) & 104(8) & 98(64) & 224(145) & 644(32) &  478(72) \\[0.5ex]
 
 $\alpha_{n,c}^{q(0)}$ &  0.112(5) & 1.5(2) & 16(1) & 35(2)	& 86(7) & 125(10) \\[0.5ex]
 
 $\alpha_{n,vc}^{q(0)}$ & 0 & 0 & 0 & $\sim0$ & $\sim0$ & $\sim0$ \\[0.5ex]
 
 Total($\alpha_{n}^{q(0)}$) & 1424(35)  & 1880(5) & 4934(107)  & 6440(246)	&  10606(736) & 8756(560)\\[0.5ex]
  
  Others & 1424~\cite{doi:10.1063/1.1578052}  & 1885~\cite{doi:10.1063/1.1578052} & 5000~\cite{doi:10.1063/1.1578052} & 6520~\cite{doi:10.1063/1.1578052} & 10470~\cite{doi:10.1063/1.1578052} & \\[0.5ex]
  
   & 1421~\cite{SAHOO2007144} & 1906~\cite{SAHOO2007144} & 4933~\cite{SAHOO2007144} & 6525~\cite{PhysRevA.83.052508}	& 10390~\cite{jiang2015effective}	 & \\[0.5ex] 
   
   & 1424~\cite{jiang2015effective}   & 1878~\cite{jiang2015effective} & 5000~\cite{jiang2015effective} & 6479~\cite{jiang2015effective} & 10521~\cite{PhysRevA.94.012505} & \\[0.5ex]  
   
   & 1420~\cite{PhysRevA.78.012515} & & 5018~\cite{PhysRevA.78.052504} & & &   \\  
 \hline
 \hline
   
	 \end{tabular}	   
	 \end{center}
	 \end{table*}
	 
\begin{table*}
\caption{\label{table2} The scalar and tensor components of the quadrupole polarizabilities (in a.u.) of the first two excited states of the Li, Na and K atoms.
Different contrubutions along with the corresponding uncertainties to these quantities are listed explicitly. The numbers in square brackets represent 
powers of 10. Our results are compared with the values reported earlier for Li.}
		\begin{center}
\begin{tabular}{|p{2.5cm}p{1.5cm}p{2.5cm}|p{2.5cm}p{1.5cm}p{2.5cm}p{2.5cm}|}
\hline
\hline
 \multicolumn{7}{|c|}{\textbf{Li}} \\
 \hline
\multicolumn{3}{|c|}{$2P_{1/2}$}  & \multicolumn{4}{c|}{$2P_{3/2}$} \\

& & & & & & \\
Contribution  & E2  & $\alpha_{n}^{q(0)}$ & Contribution  & E2 & $\alpha_{n}^{q(0)}$	& $\alpha_{n}^{q(2)}$ \\ [0.5ex]
\hline

 $\alpha_{n,v(M)}^{q(t)}$   &  &  &	 $\alpha_{n,v(M)}^{q(t)}$ &  &   &  \\[0.5ex]
   
$2P_{1/2}$	-	$2P_{3/2}$	&	24.22(2)	&	757.26(15)[5]	& $2P_{3/2}$	-	$2P_{1/2}$	&	24.22(2)	&	-378.63(75)[5]	&	378.63(75)[5]	\\[0.5ex]

$2P_{1/2}$	-	$3P_{3/2}$	&	21.147(2)	&	122.51(2)[1]	&	$2P_{3/2}$	-	$3P_{3/2}$	&	21.147(2)	&	612.6(1)[1]	&	0	\\[0.5ex]

$2P_{1/2}$	-	$4F_{5/2}$ 	&	22.99(3)	&	106.8(3)[1]	&	$2P_{3/2}$	-	$4F_{5/2}$ 	&	12.30(1)	&	152.7(2)[0]	&	109.1(7)[0]	\\[0.5ex]

$2P_{1/2}$	-	$5F_{5/2}$	&	14.45(6)	&	37.9(3)[1]	&	$2P_{3/2}$	-	$4F_{7/2}$ 	&	30.13(2)	&	917(1)[0]	&	-262.0(3)[0] \\[0.5ex]
 
  Remaining & & 461.3(1)[0] & Remaining & & 145.7(1)[1] & -776.8(6)[0]	\\[0.5ex] 
 
  $\alpha_{n,v(T)}^{q(t)}$ & & 1.01(7)[3] & $\alpha_{n,v(T)}^{q(t)}$ & & 1.00(7)[3] & -1.6(1)[2]	\\[0.5ex]
 
 $\alpha_{n,c}^{q(t)}$ & & 1.10(8)[-1] & $\alpha_{n,c}^{q(t)}$ & & 1.10(8)[-1] & 0 \\[0.5ex]
 
  $\alpha_{n,vc}^{q(t)}$ & & 0 & $\alpha_{n,vc}^{q(t)}$ & & 0 & 0 \\[0.5ex]
 
  Total($\alpha_{n}^{q(t)}$) &  & 757.31(15)[5] & Total($\alpha_{n}^{q(t)}$) & & -378.59(75)[5] & 378.62(75)[5] \\[0.5ex]
  
  Others & & 1.434[5]~\cite{PhysRevA.82.029901} &	& & & \\[0.5ex] 
 \hline
 \multicolumn{7}{|c|}{\textbf{Na}} \\
 \hline
 \multicolumn{3}{|c|}{$3P_{1/2}$}  & \multicolumn{4}{c|}{$3P_{3/2}$} \\
 $\alpha_{n,v(M)}^{q(t)}$   &  &  &	 $\alpha_{n,v(M)}^{q(t)}$ &  &   &  \\[0.5ex]
 
 $3P_{1/2}$	-	$3P_{3/2}$	&	35.939(12)	&	329.70(22)[4]	&	$3P_{3/2}$	-	$3P_{1/2}$ 	&	35.939(12)	&	-164.85(11)[4]	&	164.85(11)[4]	\\[0.5ex]

$3P_{1/2}$	-	$4P_{3/2}$	&	27.93(1)	&	257.2(2)[1]	&	$3P_{3/2}$	-	$4P_{3/2}$	&	28.008(6)	&	129.46(6)[1]	&	0 \\[0.5ex]

$3P_{1/2}$	-	$4F_{5/2}$	&	38.73(2)	&	3735(3)[0]	&	$3P_{3/2}$	-	$4F_{5/2}$ 	&	20.749(7)	&	536.4(4)[0]	&	383.2(2)[0]	\\[0.5ex]	

$3P_{1/2}$	-	$5F_{5/2}$ 	&	22.073(8)	&	106.4(1)[1]	&	 $3P_{3/2}$	-	$4F_{7/2}$ 	&	50.755(86)	&	3209(11)[0]	&	-917(3)[0]	\\[0.5ex]

Remaining & & 1085.9(4)[0]	&	Remaining & & 345.9(2)[1]	& -170.3(1)[1] \\[0.5ex]
 
$\alpha_{n,v(T)}^{q(t)}$ & & 1.62(15)[3]	&	$\alpha_{n,v(T)}^{q(t)}$ & & 1.62(14)[3]	& -2.6(2)[2] \\[0.5ex]
 
$\alpha_{n,c}^{q(t)}$ & & 1.50(12)[0]	&	$\alpha_{n,c}^{q(t)}$ & & 1.50(12)	& 0 \\[0.5ex]

  $\alpha_{n,vc}^{q(t)}$ & & 0 & $\alpha_{n,vc}^{q(t)}$ & & 0 & 0 \\[0.5ex]
 
Total($\alpha_{n}^{q(t)}$) & & 330.71(22)[4]  & Total($\alpha_{n}^{q(t)}$) & & -163.84(11)[4] & 164.60(11)[4]\\[0.5ex]
   
 \hline
  \multicolumn{7}{|c|}{\textbf{K}} \\
 \hline

\multicolumn{3}{|c|}{$4P_{1/2}$}  & \multicolumn{4}{c|}{$4P_{3/2}$} \\

 $\alpha_{n,v(M)}^{q(t)}$   &  &  &	 $\alpha_{n,v(M)}^{q(t)}$ &  &   &  \\[0.5ex]
 
$4P_{1/2}$	-	$4P_{3/2}$	&	47.08(41)	&	16.86(29)[5]	&	$4P_{3/2}$	-	$4P_{1/2}$	&	47.08(41)	&	-8.43(14)[5]	&	8.43(14)[5]\\[0.5ex]

$4P_{1/2}$	-	$5P_{3/2}$	&	33.57(22)	&	42.16(56)[2]	&	$4P_{3/2}$	-	$5P_{3/2}$	&	34.134(74)	&	219.0(1)[1]	&	0 \\[0.5ex]

$4P_{1/2}$	-	$4F_{5/2}$ 	&	55.10(98)	&	8.80(31)[3]	&	$4P_{3/2}$	-	$4F_{5/2}$	&	29.68(36)	&	12.82(31)[2]	&	9.16(22)[2] \\[0.5ex]	

$4P_{1/2}$	-	$5F_{5/2}$	&	28.46(22)	&	20.17(31)[2]	&	$4P_{3/2}$	-	$4F_{7/2}$ 	&	72.71(87)	&	7.69(18)[3]	 &	-21.98(53)[2] \\[0.5ex]

Remaining & & 18.12(10)[2]	&	Remaining & & 61.0(3)[2]	& -29.4(1)[2] \\[0.5ex]
 
 $\alpha_{n,v(T)}^{q(t)}$ & & 1.98(36)[3] &	$\alpha_{n,v(T)}^{q(t)}$ & & 1.98(36)[3] & -3.37(61)[2] \\[0.5ex]
 
$\alpha_{n,c}^{q(t)}$ & & 1.63(13)[1] &	$\alpha_{n,c}^{q(t)}$ & & 1.63(13)[1] & 0 \\[0.5ex]

  $\alpha_{n,vc}^{q(t)}$ & & 0 & $\alpha_{n,vc}^{q(t)}$ & & 0 & 0 \\[0.5ex]
 
Total($\alpha_{n}^{q(t)}$) & & 17.05(29)[5]	& Total($\alpha_{n}^{q(t)}$) & & -8.24(14)[5] & 8.38(14)[5] \\[0.5ex]
   
 \hline
 \hline

	 \end{tabular}	   
	 \end{center}
	 \end{table*}
	 
\begin{table*}
\caption{\label{table3} The scalar and tensor components of the quadrupole polarizabilities (in a.u.) of the $P_{1/2}$ and $P_{3/2}$ 
states of the Rb, Cs and Fr atoms are given. Individual contributions are listed explicitly. There are no literature values available to compare with our results.}
		\begin{center}
\begin{tabular}{|p{2cm}p{1.5cm}p{2.5cm}|p{2cm}p{1.7cm}p{2.5cm}p{2.5cm}|}
\hline
\hline
 \multicolumn{7}{|c|}{\textbf{Rb}} \\
 \hline
\multicolumn{3}{|c|}{$5P_{1/2}$}  & \multicolumn{4}{c|}{$5P_{3/2}$} \\

& & & & & & \\
Contribution  & E2  & $\alpha_{n}^{q(0)}$ & Contribution  & E2 & $\alpha_{n}^{q(0)}$	& $\alpha_{n}^{q(2)}$ \\ [0.5ex]
\hline

 $\alpha_{n,v(M)}^{q(t)}$   &  &  &	 $\alpha_{n,v(M)}^{q(t)}$ &  &   &  \\[0.5ex]
 
$5P_{1/2}$	-	$5P_{3/2}$	&	52.85(57)	&	5.16(11)[5]	&	$5P_{3/2}$	-	$5P_{1/2}$	&	52.85(57)	&	-25.80(56)[4]	&	25.80(56)[4] \\[0.5ex]

$5P_{1/2}$	-	$6P_{3/2}$	&	35.76(51)	&	5.00(14)[3]	&	$5P_{3/2}$	-	$6P_{3/2}$	&	37.47(47)	&	28.07(71)[2]	&	0 \\[0.5ex]

$5P_{1/2}$	-	$4F_{5/2}$ 	&	63.35(99)	&	12.39(39)[3]	&	$5P_{3/2}$	-	$4F_{5/2}$ 	&	34.98(51)	&	19.22(55)[2]	&	13.73(40)[2]	\\[0.5ex]	

$5P_{1/2}$	-	$5F_{5/2}$ 	&	31.10(28)	&	25.42(45)[2]	&	$5P_{3/2}$	-	$4F_{7/2}$ 	&	85.69(1.24)	&	11.53(33)[3]	&	-32.94(95)[2]	\\[0.5ex]

Remaining & & 21.56(13)[2]	&	Remaining & & 7.92(16)[3]	& -38.98(12)[2] \\[0.5ex]
 
 $\alpha_{n,v(T)}^{q(t)}$ & & 2.0(4)[3] &	$\alpha_{n,v(T)}^{q(t)}$ & & 1.99(34)[3] & -3.5(6)[2] \\[0.5ex]
 
$\alpha_{n,c}^{q(t)}$ & & 3.54(28)[1] &	$\alpha_{n,c}^{q(t)}$ & & 3.54(28)[1] & 0 \\[0.5ex]
 
  $\alpha_{n,vc}^{q(t)}$ & & -9(4)[-5] & $\alpha_{n,vc}^{q(t)}$ & & -9(4)[-5] & 0 \\[0.5ex] 
 
Total($\alpha_{n}^{q(t)}$) & & 5.40(11)[5] & Total($\alpha_{n}^{q(t)}$) & & -23.18(56)[4] & 25.18(56)[4] \\[0.5ex]
   
 \hline
 
 \multicolumn{7}{|c|}{\textbf{Cs}} \\
 \hline
 \multicolumn{3}{|c|}{$6P_{1/2}$}  & \multicolumn{4}{c|}{$6P_{3/2}$} \\

 $\alpha_{n,v(M)}^{q(t)}$   &  &  &	 $\alpha_{n,v(M)}^{q(t)}$ &  &   &  \\[0.5ex]
 
$6P_{1/2}$	-	$6P_{3/2}$	&	59.94(96)	&	28.47(91)[4]	&	$6P_{3/2}$	-	$6P_{1/2}$	&	59.94(96)	&	-14.23(46)[4]	&	14.23(46)[4] \\[0.5ex]

$6P_{1/2}$	-	$7P_{3/2}$	&	37.2(9)	&	5.63(27)[3]	&	$6P_{3/2}$	-	$7P_{3/2}$	&	41.81(81)	&	3.76(15)[3]	&	0 \\[0.5ex]	

$6P_{1/2}$	-	$4F_{5/2}$	&	73.34(1.71)	&	17.76(83)[3]	&	$6P_{3/2}$	-	$4F_{5/2}$ 	&	42.30(86)	&	3.08(12)[3]	&	22.01(89)[2]	\\[0.5ex]

$6P_{1/2}$	-	$5F_{5/2}$ 	&	33.68(39)	&	31.5(8)[2]	&	$6P_{3/2}$	-	$4F_{7/2}$	&	103.61(2.09)	&	18.49(75)[3]	&	-5.28(21)[3]	\\[0.5ex]

Remaining & & 24.80(13)[2]	&	Remaining & & 10.61(16)[3]	& -55.28(18)[2] \\[0.5ex]
 
 $\alpha_{n,v(T)}^{q(t)}$ &  & 1.9(5)[3] &	$\alpha_{n,v(T)}^{q(t)}$ & & 1.8(4)[3] & -3.5(9)[2] \\[0.5ex]
 
$\alpha_{n,c}^{q(t)}$ & & 8.64(69)[1] &	$\alpha_{n,c}^{q(t)}$ & & 8.64(69)[1] & 0\\[0.5ex]

  $\alpha_{n,vc}^{q(t)}$ & & -5(2)[-4] & $\alpha_{n,vc}^{q(t)}$ & & -5(2)[-4] & 0 \\[0.5ex] 
 
Total($\alpha_{n}^{q(t)}$) & & 31.57(92)[4] & Total($\alpha_{n}^{q(t)}$) & & -10.45(47)[4] & 13.34(46)[4] \\[0.5ex]
   
 \hline
  \multicolumn{7}{|c|}{\textbf{Fr}} \\
 \hline
 \multicolumn{3}{|c|}{$7P_{1/2}$}  & \multicolumn{4}{c|}{$7P_{3/2}$} \\

 $\alpha_{n,v(M)}^{q(t)}$   &  &  &	 $\alpha_{n,v(M)}^{q(t)}$ &  &   &  \\[0.5ex]
 
$7P_{1/2}$	-	$7P_{3/2}$	&	61.25(1.46)	&	9.76(47)[4]	&	$7P_{3/2}$	-	$7P_{1/2}$	&	61.25(1.46)	&	-4.88(23)[4]	&	4.88(23)[4] \\[0.5ex]

$7P_{1/2}$	-	$8P_{3/2}$	&	30.66(1.3)	&	3.61(31)[3]	&	$7P_{3/2}$	-	$8P_{3/2}$	&	44.90(1.2)	&	4.54(25)[3]	&	0 \\[0.5ex]

$7P_{1/2}$	-	$5F_{5/2}$ 	&	70.96(2.7)	&	1.59(12)[4]	&		$7P_{3/2}$	-	$5F_{5/2}$	&	47.65(1.4)	&	4.09(24)[3]	&	2.92(17)[3]	\\[0.5ex]

$7P_{1/2}$	-	$6F_{5/2}$ 	&	32.84(53)	&	28.80(94)[2]	&	$7P_{3/2}$	-	$5F_{7/2}$	&	116.75(3.44)	&	2.46(14)[4]	&	-7.0(4)[3]	\\[0.5ex]

Remaining & & 19.61(33)[2]	&	Remaining & & 13.43(31)[3]	& -88.66(19)[2] \\[0.5ex]
 
 $\alpha_{n,v(T)}^{q(t)}$ & & 2.20(88)[3] &	$\alpha_{n,v(T)}^{q(t)}$ & & 1.91(57)[3]  & -3.6(1.1)[2] \\[0.5ex]
 
$\alpha_{n,c}^{q(t)}$ & & 1.25(10)[2] &	$\alpha_{n,c}^{q(t)}$ & & 1.25(10)[2] &	0 \\[0.5ex]
 
   $\alpha_{n,vc}^{q(t)}$ & & -3(2)[-4] & $\alpha_{n,vc}^{q(t)}$ & & -3(2)[-4] & 0 \\[0.5ex] 
   
Total($\alpha_{n}^{q(t)}$) & & 12.43(49)[5] & Total($\alpha_{n}^{q(t)}$) & & -1468(2844)[-1] & 3.63(24)[4] \\[0.5ex]
   
\hline
\hline   
   
	 \end{tabular}	   
	 \end{center}
	 \end{table*}

\section{Results and discussion}\label{sec4}

\subsection{Quadrupole polarizability of ground state}\label{sec4a}

We present the static values of $\alpha_{n}^{q(0)}$ of the ground states of alkali-metal atoms and compare them with other available 
data in Table~\ref{table1}. 
The scaled SD values of matrix elements for the main part of the polarizability have been taken as final values as recommended in previous studies for E2 transitions~\cite{PhysRevA.78.022514,PhysRevA.78.052504,PhysRevA.95.042507}. 
The breakdown of total polarizability into the main, tail, core and valence-core polarizabilities are presented.
The valence-core contributions for Li, Na and K are zero due to non-availability of $D$ orbitals in the core of these atoms whereas very 
insignificant contributions have been encountered for Rb, Cs and Fr. To provide estimates for error bars in the net value of each contribution of 
polarizability, we have incorporated the uncertainties for main, tail and core using different procedures. The uncertainty in the main part of 
valence polarizability is solely attributed to the uncertainty in matrix elements of the dominant transitions. The percentage uncertainty in 
tail part has been  estimated by calculating the percentage deviation between the polarizability contribution of highest lying transition of main
part calculated by DF and SD method. Recent experimental measurement on quadrupole core polarizability by Berl \textit{et al.}
\cite{PhysRevA.102.062818} are found to be in good agreement with the core polarizability calculated using RPA for Rb. However, the RPA value for 
Rb gives maximum of 8\% uncertainty when compared with the experimental value. Therefore, we have assigned 8\% uncertainty to the core 
polarizability for all the atoms. 
The net uncertainty in the total value of polarizability has been accomodated by adding individual uncertainties in quadrature.
 
As Table~\ref{table1} suggests, the main part of valence polarizability is responsible for over 90$\%$ of the total polarizability value for every considered atom.
We ascribe 3\%, 5\%, 65\%, 65\%, 5\% and 15\% uncertainty to the tail part for Li, Na, K, Rb, Cs and Fr, respectively. Our values for the static quadrupole polarizability of Li, Na and K are found to 
be 1424(35), 1880(5) and 4934(107) a.u., respectively. Our resulted values match very well with other
theoretical values calculated using semi-empirical~\cite{jiang2015effective}, RMBPT~\cite{doi:10.1063/1.1578052} and CCSD~\cite{SAHOO2007144} methods. For K, the quadrupole polarizability value recommended by Safronova \textit{et al.} is 5018 using the SD values with 70 splines~\cite{PhysRevA.78.052504}. 
To authenticate our precisely calculated E2 matrix elements of the dominant transitions, we compare our E2 matrix element for Rb, Cs and Fr with the values that are available in the literature. 
Our E2 matrix 
elements , 32.88(74) and 40.29(90) a.u., from the $5S_{1/2} \rightarrow 4D_{3/2}$ and $5S_{1/2} \rightarrow 4D_{5/2}$ transitions, respectively of Rb are in excellent agreement with the values of 32.94(14) and 40.37(17) a.u. that are recommended by 
Safronova \textit{et al.}~\cite{PhysRevA.95.042507}. Furthermore, Gossel \textit{et al.} reported the matrix element of 33.42 a.u. for 
the $5S_{1/2} \rightarrow 4D_{3/2}$ transition calculated using the relativistic Hartree-Fock approximation in a $V^{N-1}$ potential 
\cite{PhysRevA.88.034501} which lies within the uncertainty limit of our value. Theoretical 
E2 values of the corresponding matrix elements for Cs, 33.61(28) and 41.46(24) a.u., for  the $6S_{1/2} \rightarrow 5D_{3/2}$ and $6S_{1/2} \rightarrow 5D_{5/2}$ transitions, respectively are in excellent agreement with our values, 33.62(1.77) and 41.56(2.07) a.u. as reported in a 
recent study~\cite{PhysRevA.95.042507}. These 
values for the $6S_{1/2} \rightarrow 5D_{3/2}$ transition computed using the highly accurate methods deviate from the experimental value which has 
been measured by the method of two-photon ionization of the ground $6S$ state, using the $5D$ as an intermediate state~\cite{GLAB1981262} by 
2$\%$ only.
On comparing the E2 matrix elements of most 
dominant transitions of Fr \textit{i.e.}, $7P_{1/2} \rightarrow 6D_{3/2}$ and $7P_{1/2} \rightarrow 6D_{5/2}$, our values, 33.40(1.33) and 41.54(1.47) a.u. are again in reasonable agreement with the values, 33.43(19) and 
41.58(18) a.u., recommended by Safronova \textit{et al.}~\cite{PhysRevA.95.042507}.
Our final quadrupole polarizability values of 6440(246) and 10606(736) a.u. of Rb and Cs, respectively advocate the results 
evaluated by Safronova \textit{et al.}~\cite{PhysRevA.83.052508,PhysRevA.94.012505} and are comparable to 
the values calculated using RMBPT~\cite{doi:10.1063/1.1578052} and semi-empirical~\cite{jiang2015effective} approaches.
Combining all the individual contributions for Fr, the ground state quadrupole polarizability value comes out to be 8756(560) a.u.. The trend of rising 
quadrupole polarizability down the group I breaks at Cs as Fr offers lower value of the ground state quadrupole polarizability than Cs. 
Quantitatively, both the matrix elements and energies of the transitions play principal roles in the determination of this lower value of 
polarizability. The smaller values of the E2 matrix elements and large values of the energy differences among the primary transitions of Fr are 
responsible for such smaller value as compared to its preceding alkali atom. The quadrupole polarizability for Fr has not been explored to date 
by any other group. Nevertheless, the accuracy in the ground state quadrupole polarizability values of all other alkali-metal atoms makes the 
resulted polarizability value for Fr as much authentic as for other considered alkali-metal atoms.

\subsection{Quadrupole polarizabilities of excited states}\label{sec4b}

The calculated values of static quadrupole polarizabilities of the first two excited states, $nP_{1/2}$ and $nP_{3/2}$, of alkali-metal atoms are 
presented in Tables~\ref{table2} and \ref{table3}. Same procedures have been followed for the calculations of these quantities, i.e.
main, tail, valence-core and core contributions as discussed in Sec.~\ref{sec4a}. The $nP_{3/2}$ state quadrupole polarizabilities have 
contributions from the scalar as well as tensor components. 
The main contributions arising from the most dominant E2 matrix elements are quoted explicitly here, while the rest are given separately as
`Remaining' in the above tables. The difference between the tail and `Remaining' contributions is that the tail contributions are coming 
from the high-lying states including continuum and estimated using the DF method, while the `Remaining' contributions are arising from the 
low-lying bound states and estimated more accurately by combining the E2 matrix element from the scaled SD methods and the experimental 
energies. We could not find any other values in the literature for comparative analysis of our quadrupole values apart from the result for the $2P_{1/2}$ state
of Li as discussed below.

\subsubsection{Li}

Table~\ref{table2} consists of static quadrupole polarizability values for both the $2P_{1/2}$ and $2P_{3/2}$ states of Li along with the main,
tail, and core contributions. For Li, we used energies for the $(6-8)F_{5/2,7/2}$ states from the SD method as the NIST energies are not available.
For the main part, other than the listed transitions, $2P_{1/2} \rightarrow  (4-7)P_{3/2}, (6-8)F_{5/2}$ transitions have been included, the 
contribution of these transitions are given in Remaining part of $\alpha_{n,v(M)}^{q(t)}$.
As shown in Table~\ref{table2}, the individual contributions from the dominant transitions considered for the 
$2P_{1/2}$ state of Li clearly suggest that the largest contribution towards the total polarizability value is coming from the $2P_{1/2} \rightarrow 2P_{3/2}$ 
transition by the reason of large matrix element as well as a very small difference between the experimental excitations energies (0.34 cm$^{-1}$) of the 
corresponding states. This small difference in the energy can be attributed to very small fine splitting of the $2P$ state coming into effect due 
to spin-orbit coupling. Another effective contribution towards the main part of total polarizability of the $2P_{1/2}$ state has been provided by 
the $2P_{1/2} \rightarrow 4F_{5/2}$ transition. For the $2P_{1/2}$ state, the tail part offers a very little contribution ($<1\%$) whereas the
core polarizability is 0.11 a.u.. A7\% uncertainty has been considered in the tail part. Total polarizability value of the $2P_{1/2}$ state of
Li is found to be $757.31(15) \times 10^5$ a.u. We found another work reported by Wansbeek \textit{et al.} for the calculation of quadrupole polarizability 
of the 2$P_{1/2}$ excited state of Li atom using \textit{ab initio} CCSD(T) method~\cite{PhysRevA.78.012515,PhysRevA.82.029901}. As can be 
noticed from the Table~\ref{table2}, there is a large deviation in the results between this work and the value reported in Ref. \cite{PhysRevA.78.012515,PhysRevA.82.029901}.
From the contributions explicitly quoted in Table~\ref{table2}, it is obvious that such a huge deviation would have been caused due to estimate 
of different contribution from its fine-structure partner 2$P_{3/2}$ excited state as the magnitudes of other contributions are relatively small. A furthermore analysis suggests that the fine-structure 
splitting of the $2P$ state is extremely small and it is a challenge to estimate this splitting as precisely as the experimental value using a 
numerical calculation without considering contributions from the higher-order relativistic effects. This is the only reason why we observe a huge difference 
between the {\it ab initio} calculation of Ref. \cite{PhysRevA.78.012515,PhysRevA.82.029901} and the present work, where we have considered the
experimental energies in the sum-over-states approach to determinate the quadrupole polarizabilities. From this view point, the result reported 
in this work is more reliable. For the $P_{3/2}$ state, we have estimated contributions from a large number of dominant transitions \textit{i.e.,} $2P_{3/2} 
\rightarrow (2-7)P_{1/2}, (3-8)P_{3/2}, (4-8)F_{5/2,7/2}$ out of which 4 dominant transitions are listed in the table, in the sum-over-states 
approach for the evaluation of quadrupole polarizatity. The transitions which have 
not been explicitly mentioned are included in the Remaining part of the main valence contribution. It can be observed that the contribution due to the first transition in Table~\ref{table2} towards the scalar 
component of the main part of the total polarizability is negative, which is ascribed to lower energy value of the $2P_{1/2}$ level than the 
$2P_{3/2}$ level. However, for tensor component of the $2P_{3/2}$ state, the same transition provides a positive contribution which is attributed 
to negative $W_{n,k}^{q(2)}$ coefficient of the tensor component that negates with negative sign in Eq. (\ref{tensor}). Moreover, the contributions of 
$P_{3/2} \rightarrow P_{3/2}$ transitions for any principal quantum number in the main part of the tensor polarizability for $P_{3/2}$ states of 
all the alkali-metal atoms are zero because 6-j symbol in Eq. (\ref{eq9}) vanish when triangle conditions are not fulfilled. Owing to negative 
Wigner coefficient $W_{n,k}^{q(2)}$ for the $2P_{3/2} \rightarrow mF_{7/2}$ transitions, where $m>4$, the contributions of these 
transitions are negative for main part of the tensor component of the polarizability. This type of behavior is true for all the alkali-metal
atoms. Since a lot of dominant transitions have been examined for the main part of tensor component, a very small percentage of the tail part has 
been encountered giving scalar value of quadrupole polarizability of $-378.59(75) \times 10^5$ a.u. and the tensor polarizability value as  
$378.62(75) \times 10^5$ a.u..

\subsubsection{Na}

Table~\ref{table2} provides the individual contributions of the main, tail and core of the total quadrupole polarizability for the $3P_{1/2}$ and 
$3P_{3/2}$ states of Na. Using the experimental energies and the precisely calculated E2 matrix elements of all the dominant transitions $3P_{1/2} 
\rightarrow (3-8)P_{3/2},(4-8)F_{5/2}$, the value of the main part for the $3P_{1/2}$ state is amounting about 95\% contribution towards the 
total polarizability value. Such large fraction is solely attributed to the contribution of 
the first transition of the main polarizability given in the table. The tail and core contribution are quite small with 9\% and 8\% uncertainties, respectively. With all the contributions
of polarizability, the total value of the $3P_{1/2}$ state comes out to be $330.71(22)\times 10^4$. Same can be noticed for scalar and tensor components of the
quadrupole polarizability of the $3P_{3/2}$ state for which the $3P_{3/2} \rightarrow 3P_{1/2}$ transition is giving dominant contribution. 
Other than listed transitions, we included the contributions from $3P_{3/2} \rightarrow (4-8)P_{1/2},(5-9)P_{3/2},(5-8)F_{5/2,7/2}$ in the 
Remaining part of the main valence contribution. After adding all the individual contributions and uncertainties in 
the quadrature, the total polarizability values of the scalar and tensor components for $3P_{3/2}$ state are $-163.84(11)\times 10^4$ and $164.60(11)\times 10^4$ 
a.u., respectively.

\subsubsection{K}

We present all the contributions to the quadrupole polarizability for the $4P_{1/2}$ and $4P_{3/2}$ excited states of K in Table~\ref{table2}.
Around 98\% of the share of total polarizability has been imparted by the main part of the valence polarizability which include contributions from $4P_{1/2} \rightarrow (4-9)P_{3/2},(4-8)F_{5/2}$ transitions. Remainder share is 
coming from both the tail and core polarizabilities for the $4P_{1/2}$ state. The 8\% and 18\% uncertainty has been given to the core and tail 
polarizabilities respectively. Net quadrupole polarizability value of $4P_{1/2}$ state of K comes out to be $17.05(29)\times 10^5$ a.u..
For the $4P_{3/2}$ state, one can observe that the large matrix elements are rendered by the $4P_{3/2} \rightarrow 4P_{1/2}, 5P_{3/2}, 4F_{5/2}$ and 
$4F_{7/2}$ transitions. The largest matrix element given by the $4P_{3/2} \rightarrow 4F_{7/2}$ transition does not provide immense 
contribution towards total polarizability due to significant difference in the energy state. Other transitions ($4P_{3/2} \rightarrow (5-9)P_{1/2},
(6-10)P_{3/2}, (5-8)F_{5/2,7/2}$) which account for very less contribution as compared to the dominant ones have been listed as Remaining in
the table. From the DF method, tail part has been estimated with 18\% uncertainty. Adding all the contributions, the scalar and tensor polarizabilities are $-8.24(14)\times 10^5$ a.u. and $8.38(14)\times 10^5$ a.u., respectively for $4P_{3/2}$ state.

\subsubsection{Rb} 

The total polarizability values of the $5P_{1/2}$ and $5P_{3/2}$ excited states for Rb are given in Table~\ref{table3} with individual 
contributions from the main, tail, core and valence-core correlations. For the $5P_{1/2}$ state, we carried out precise E2 matrix element 
calculations of the $5P_{1/2} \rightarrow (5-10)P_{3/2},(4-5)F_{5/2}$ transitions. The $5P_{1/2} \rightarrow 5P_{3/2}$ transition is offering an overwhelming 
contribution of around $96\%$ to the total polarizability for the $5P_{1/2}$ state. The tail and core contributions are only 0.4\% and 0.006\%, 
respectively, of the total polarizability for the $5P_{1/2}$ state with tail uncertainty of 22\% whereas valence-core contribution is
nearly zero. Thus, a total quadrupole polarizability of $5.40(11)\times 10^5$ a.u. has been encountered for the $5P_{1/2}$ state. For 
the $5P_{3/2}$ state of Rb, similar findings can be observed for both the scalar and tensor components. The  
$5P_{3/2} \rightarrow (5-10)P_{1/2},(6-11)P_{3/2}, (4-8)F_{5/2,7/2}$ transtions have been considered for the main part out of which contributions 
from the 4 transitions are listed in the above table and the contributions for the remaining have been listed as Remaining. The tail contribution is very small in 
comparison to the main part. We assign maximum 17\% uncertainty to the tail contribution for both the scalar and tensor components. The scalar and 
tensor quadrupole polarizabilities of the $5P_{3/2}$ state are $-23.18(56)\times 10^4$ and $25.18(56)\times 10^4$ a.u., respectively.

\subsubsection{Cs} 

The matrix elements and polarizability contributions from the considered transitions of the main part for Cs have been summarized in 
Table~\ref{table3}. The first transition, i.e. $6P_{1/2} \rightarrow 6P_{3/2}$, of the main part in the given table for the $6P_{1/2}$ state is 
accountable for large value of total polarizability. Other tabulated transitions are also contributing dominantly. Remaining contributions of the 
main part include contributions from the $6P_{1/2}\rightarrow (7-11)P_{3/2},(5-8)F_{5/2}$ tranistions. The tail and core correlations provide $\sim 0.6\%$ and $\sim 0.027\%$  contributions of the 
total polarizability value. 27\% uncertainty has been assigned to the tail part. Thus, the total polarizability value of the $6P_{1/2}$ state 
for Cs turns out to be $31.57(92)\times 10^4$ a.u.. For the scalar  component of the $6P_{3/2}$ state, the assertive positive 
contributions in the main part are given by the $6P_{3/2} \rightarrow (7-11)P_{1/2},(7-12)P_{3/2},(4-8)F_{5/2,7/2}$ transitions, core and tail 
part cancel with some share of the negative contribution given by the most prominent  $6P_{3/2} \rightarrow 6P_{1/2}$ transition, ultimately 
giving a small value for the main polarizability of $-10.64 \times 10^4$, whereas tensor polarizability is $13.37 \times 10^4$ a.u.. The assigned 
tail uncertainty for the $6P_{3/2}$ state is 25\%. Finally, the total polarizability values for scalar and tensor components of the $6P_{3/2}$ 
state are $-10.45(47)\times 10^4$ and $13.34(46) \times 10^4$, respectively.

\subsubsection{Fr} 

The total polarizability value of the $7P_{1/2}$ and $7P_{3/2}$ states for Fr with the main, tail, core and valence-core contributions is given 
in Table~\ref{table3}. We calculated the E2 matrix elements for the $7P_{1/2} \rightarrow (7-12)P_{3/2}, (4-8)F_{5/2}$ transitions to estimate 
the quadrupole polarizability of the $7P_{1/2}$ state and the E2 matrix elements for the $7P_{3/2} \rightarrow (7-12)P_{1/2}, (8-13)P_{3/2} 
(4-8)F_{5/2,7/2}$ transitions of determining quadrupole polarizability of the $7P_{3/2}$ state. For the $(5-8)F_{5/2,7/2}$ states, we have used our SD excitation energy values as the energies from the NIST database are not available. 
Our energy values for the $(5-8)F_{5/2,7/2}$ states agree well with values recommended by Tang \textit{et al.} calculated using relativistic Fock 
space multi-reference coupled-cluster method~\cite{PhysRevA.96.022513}. The percentage differences of our energy values with respect to the values 
evaluated by Tang \textit{et al.} for the $(5-8)F_{5/2,7/2}$ states ranges from 0.04 to 0.09\%. For Fr, the $7P_{1/2} \rightarrow 7P_{3/2}$ and $7P_{1/2} \rightarrow 5F_{5/2}$ transitions play major roles in total polarizability value of the $7P_{1/2}$ state as can be observed from the individual contribution 
given in Table~\ref{table3}. Though matrix elements are large, the effect of polarizability contribution is not prodigious. The reason behind this
being the large doublet separation of the $7P$ state. Including the tail and core polarizabilities of $2.20(88)\times 10^3$ a.u. and 125 
a.u. with the corresponding uncertainties of 40\% and 8\%, the total polarizability value for the $7P_{1/2}$ state is $12.43(49) \times 10^4$ a.u..
For the $7P_{3/2}$ state, the $7P_{3/2} \rightarrow 7P_{1/2}, 8P_{3/2}, 5F_{5/2,7/2}$ transitions for scalar component contribute dominantly.
It is worth to bring to notice that unlike all other atoms considered in this work, the main part of scalar component for the $7P_{3/2}$ state 
is giving very small negative contribution. With tail part having 30\% uncertainity and core contribution, the final value for the scalar component for the $7P_{3/2}$ state is -146.8 a.u. with a 
large uncertainty of 2844 a.u.. For tensor component, the $7P_{3/2} \rightarrow 7P_{1/2}, 8P_{1/2}, 5F_{5/2,7/2}$ transitions contribute 
dominantly for the main part leading to total polarizability of $3.63(24)\times 10^4$ a.u., respectively after adding all other components.
However, the total polarizability value coming from both the scalar and tensor components of the $7P_{3/2}$ state depend upon the magnetic 
sublevels $M_{J}$ values.  

\section{Conclusion}\label{sec5}

We have presented the static quadrupole polarizabilities of the ground state and the first two excited states $nP_{1/2,3/2}$ of the 
alkali-metal atoms. Uncertainties to these quantities are reduced by using very precise values of electric quadrupole matrix elements 
of a large number of intermediate states and considering experimental energies. The electric quadrupole matrix elements were evaluated 
by employing an all-order relativistic many-body method in the singles-doubles scaling procedure that takes experimental correlation
effects into account. The calculated quadrupole polarizability values were
validated by reproducing the values for the ground states of the above atoms with the literature values. This confirms the credibility of 
our results for the excited states. To understand their accuracies further, breakdown of contributions towards the net values along with the 
quadrupole matrix elements and their uncertainties for dominant transitions are also given. 
The precise values of quadrupole matrix elements given in this work can be used to estimate the dynamic quadrupole polarizabilities of the 
considered states at real and imaginary frequencies, which are useful for many applications. The reported quadrupole polarizability values can be helpful for estimating 
systematics associated with the high-precision experiments using alkali-metal atoms. 

 \section{Acknowledgements}
The work of B. A. is supported by SERB-TARE research grant no. TAR/2020/000189, New Delhi, India. The employed all order method was developed in the group of Professor M. S. Safronova of the University of Delaware, USA.


\end{document}